\documentclass[a4paper,12pt]{article}
\usepackage{amssymb,amsmath,pxfonts,latexsym} 
\usepackage{wasysym} 
\usepackage[T1]{fontenc}   
\usepackage[english]{babel}
\usepackage{xcolor}
\definecolor{hypercolor}{rgb}{0,0.2,0.7}

\usepackage[breaklinks,final]{hyperref}
\hypersetup{
  colorlinks,
  linkcolor=hypercolor,
  urlcolor=hypercolor,
  citecolor=hypercolor,
  bookmarksnumbered,
  bookmarksdepth=2,
  pdfauthor={Ivan Zhogin},
  pdftitle={Absolute Parallelism: Spherical symmetry and  Singularities}
}

\def\a{\alpha}          \def\b{\beta}            \def\g{\gamma}
\def\d{\delta}               \def\ve{\varepsilon}
                    \def\l{\lambda}
\def\m{\mu}             \def\n{\nu}              
            \def\s{\sigma}           \def\t{\tau}
      \def\om{\omega}          \def\vk{\varkappa}
          \def\D{\Delta}           \def\L{\Lambda}

\def\be{\begin{equation}}            \def\ee{\end{equation}}
\def\ba#1{\begin{array}{#1}}         \def\ea{\end{array}}

\def\fr#1#2{\textstyle\frac{#1}{#2}}
\def\bA{{\mathbf A}} \def\bB{{\mathbf B}}
\def\cL{\mathcal{L}}
 
\def\bH{\mathbf H} \def\bG{\mathbf G}

\begin{document}

\renewcommand{\refname}{\bf\large LITERATURE CITED}
\pagestyle{empty}   
\centerline{\Large\bf Absolute Parallelism: spherical symmetry and singularities}  

\vspace{2.5mm}

\hspace*{11.5mm} {\bf I.\,L.\ Zhogin} \ (zhogin at mail.ru) 
\hspace{55mm} UDC  530.12:531.51
\\[-.1 mm]   

\hspace*{1.5mm}
\begin{minipage}[t]{0.91\textwidth}
A one-parameter class of compatible second order equations of frame field 
$h^{\,a}{\!}_\mu(x^\nu)$ 
is considered, characterized by the fact that after the substitution
$h^{\,a}{\!}_\mu = H^p H^a{}_\mu$  (where $H = \det H^a{}_\mu$,  $p$ is the constant depending on the 
dimensionality $D$ and on the equation parameter $\sigma$), 
the equations can be written in a
tri-linear form which contains the matrix $H_a{}^\mu$ and its derivatives, 
but not $H^a{}_\mu$. The equations remain regular for degenerated matrices 
$H_a{}^\mu$ if $r = \textrm{rank}\, H_a{}^\mu \ge 2$, the points with $r< D$ being (generally) singular.

 It is shown that, with one coordinate choice (gauge), a single second order equation remains and there exist spherically symmetric solutions with arising singularities ($r< D$).
On the other hand, a more reasonable (covariant) choice of the radius and time reduces the problem to a system of two first order equations looking like Chaplygin gas dynamics,  where (non-stationary) $O_{D-1}$-solutions are seemingly free of emerging singularities.{}\footnote{\ 
Translated from Izvestiya Vysshikh Uchebnykh Zavedenii, Fizika, No.~9, pp.~47--52, September, 1991. Original article submitted November 16, 1990.}   

\end{minipage} 
 \\[1.5mm]

 1. While being the simplest generalization of General Relativity, 
the theory of Absolute Parallelism (AP)
 \cite{EiMa} possesses much more diverse features and tools than the former does. Apart from the metric, there is a sort of
 ''electromagnetic field'', and there is a possibility of having particle-like solutions -- 
 topological quasi-solitons \cite{me_T}. The most urgent problem of general relativity is, probably, 
 that of singularities \cite{HawEll, LL}. 
 This is why it seems interesting to find out what are the singularities
 in the AP theory -- especially in those cases which allow 
for a tri-linear form of equations 
 \cite{me_3}.

 Let us consider first the classification of 
 Einstein-Mayer \cite{EiMa} of compatible systems of equations
 of second order in the field $h^a{}_\mu$ (selecting tri-linear ones). 
 Aside from being generally covariant, the AP equations are invariant under 
 the global transformations (scalings and rotations)
\be \label{e1}
 {}^*h^a{}_\mu (x^\nu) = \kappa \, s^a{}_b \, h^b{}_\mu(x^\nu), 
 \textrm{ where } \kappa >0,\, s^a{}_b \in O(1,D-1); \ \kappa,s^a{}_b= \textrm{const} \, .
\ee
\noindent
We introduce the metric 
($\m,\n = 0, 1, \ldots, D-1; \ a,b = \underline{0},\underline{1}, \ldots, \underline{D-1}$)
\[ g_{\mu\nu} = \eta_{ab} h^a{}_\mu h^b{}_\nu =h^a{}_\mu h_{a\nu},
 \textrm{ where } 
 \eta_{ab} = \textrm{diag}(-1,1,\ldots,1)= \eta^{ab}, 
\]
and covariant differentiation with the usual symmetric Levi-Civita connection.
Under coordinate transformations, the Roman indices are of the scalar type, and the Greek ones of the vector type.

When writing manifestly covariant equations, we omit the matrices $\eta^{ab}$ and $g^{\mu\nu}$ in contractions, since $g^{\mu\nu}{}_{;\l} \equiv 0$.
The index type can be changed by means of $h_{a\mu}$, \,
e.g.\ $\L_{\mu\nu\l} = h_{a\mu}\L_{a\nu\l}$. 

We also use the next notations and definitions
 (see details in \cite{me_T}; as usual, $U_{\Gamma,\mu}$ means $\partial U_\Gamma /\partial x^\mu$):
\be \label{e2}
\L_{a\mu\nu} =h_{a\mu;\nu} -h_{a\nu;\mu} = h_{a\mu,\nu} -h_{a\nu,\mu}
 = 2 h_{a[\mu;\nu]}  ; \  \ \Phi_\mu = \L_{\nu\nu\mu};
\ \ f_{\mu\nu} = \Phi_{\mu;\nu} - \Phi_{\nu;\mu} \, ;
\ee
\be \label{e3}
S_{\mu\nu\l} =\L_{\mu\nu\l} + \L_{\nu\l\mu } + \L_{\l\mu\nu} = 
\L_{\mu\nu\l} + (\mu\nu\l) = 3\L_{[\mu\nu\l]}; \,
\ \textrm{note, } \,  
h_{a\m} h_{a\n;\l} =\fr12 S_{\m\n\l} - \L_{\l\m\n}\,.
\ee 
\noindent
The definition of the tensor $\L_{a\m\n}$ in \eqref{e2} results in the evident identity and its contraction:
\be \label{id}
\L_{a[\mu\nu;\l]} \equiv 0 \ (\textrm{and } f_{[\m\n;\l]}\equiv 0),
  \textrm{ or }
\L_{\ve\mu\nu;\l} + (\fr12 S_{\l\ve\tau} - \L_{\l\ve\tau})\L_{\tau\mu\nu} +
 (\mu\nu \l) \equiv 0;
\ee
 \be \label{id2}
\L_{\l\mu\nu;\l} + f_{\m\n} + \L_{\ve\tau [\mu}\L_{\nu] \ve\tau}
-\Phi_\l \L_{\l\m\n} \equiv 0\, ,
\ \textrm{\,or } \, \L_{abc,a} + f_{bc} \equiv 0.
\ee

2. To analyze the AP equations, it is convenient to separate the symmetric and skew-symmetric parts 
[see Eqs.~\eqref{e2}--\eqref{id2}; $\s, \; \t,\; \a_i$ are some parameters, constants]:
\be \label{e6S}
{\bG}_{\mu\nu} =  \L_{\mu\nu\l;\l} +  \L_{\nu\mu\l;\l} +
\s (\Phi_{\m;\n} + \Phi_{\n;\m} - 2 g_{\m\n} \Phi_{\l;\l}) + Q_{\m\n}(\L^2)=0;
\ee
\be \label{e7A}
{\bH}_{\mu\nu} = S_{\mu\nu\l;\l} +
\t f_{\m\n} + a_1 S_{\m\n\l}\Phi_\l + a_2 \L_{\m\n\l}\Phi_\l +
a_3(\L_{\m a b}\L_{ab\n} - \L_{\n a b}\L_{ab\m}) = 0;
\ee
Equations \eqref{e7A} contains all possible terms quadratic in $\L$; 
$Q_{\m\n} = Q_{\n\m}$ -- a symmetric tensor. 

Differentiating Eqs.~\eqref{e6S} and \eqref{e7A}, we obtain 
two "Maxwell-like equations":
\[ \bG_{\m\n;\n} = (\s -1) [f_{\m\n;\n} - J^{(1)}_\m(\L\L')] = 0; \
\bH_{\m\n;\n} = \t [f_{\m\n;\n} - J^{(2)}_\m(\L\L')] = 0.
\]
In order that the system \eqref{e6S}, \eqref{e7A} be compatible, 
it is necessary (and sufficient if $\t \neq 0$) that 
the following equation becomes identity (we exclude the principal terms,
$\Phi''$):
\be \label{id8}
\t \bG_{\m\n;\n} + (1-\s) \bH_{\m\n;\n} = 0 \
 \textrm{(or } J^{(1)}_\m - J^{(2)}_\m \equiv 0).
\ee
The next identity [for the prolonged equation $f_{\m\n;\n}= J_\m(\L\L')$], $J_{\m;\m}\equiv 0$, is automatically valid, 
if $\t \neq 0$, because $J^{(2)}_\m$ is a divergence 
of a skew-symmetric tensor.

The conditions necessary for "tri-linearity" \cite{me_3}
have the following form:
\be \label{tri-c}
\t= 1-3\s=-a_1, \ a_2 =a_3 = 0.
\ee

If $a_2=a_3 = 0$, there exist solutions with $f_{\m\n}=0$ \cite{me_3}.

Let us also consider the possibility of adding to the system 
\eqref{e6S}, \eqref{e7A} the equation $S_{\m\n\l}=0$.
If  in Eq.~\eqref{e7A} $a_2=0$, then the equation 
$\bH_{[\m\n;\l]} = 0$  becomes identity. 
However, there is also the identity \eqref{id} to take into account.
The totally antisymmetric part of Eq.~\eqref{id} 
(loses its principal terms and) results in the next nonregular
first order equation:
$  \L_{\a[\ve\l} \L^\a{}_{\m\n]}  =0.
$

This nonregular equation becomes identity if $D\leq 3$ 
or if there is a high enough 
symmetry so there are no ways in constructing an antisymmetric tensor 
of the rank four. 

3. Out of the four classes of compatible [i.e., satisfying the identity 
\eqref{id8}] systems of equations found by Einstein and Mayer
in \cite{EiMa}, the simplest is the
class II$_{22112}$:
\be \label{eA}
\bA_{a\m} = (\L_{a\m\n} + \g S_{a\m\n} - \s h_{a\m}\Phi_\n +
\s h_{a\n}\Phi_\m )_{;\nu} =0, \ \bA_{a\m;\m} \equiv 0. 
\ee
In Eq.~\eqref{eA}, $\t = (1-\s)/(1 + 2\g)$; 
the conditions \eqref{tri-c} are satisfied if 
$\g = -1,\, \s = 1/2$.\footnote{For the simplest case 
$\g = 0= \s $ the linearized Eqs.~\eqref{eA} [together with the linearized
identity (\ref{id})] look like $D$-fold Maxwell systems, so the number of
degrees of freedom equals $D(D-2)$. This is valid for all cases excepting some "pathologies" where compatibility (and regularity) of equations is lost 
(see further comments).}

The class I$_{12}$ (in \cite{EiMa}) includes compatible equations which follow from the variational principle. Let us consider the Lagrangian 
(density; $h=\det h^a{}_\m$) 
$h{\cal L}$:
\[  {\cL} = \fr14 \L_{abc}\L_{abc} + \fr\g{12} S_{abc} S_{abc}
- \fr\s2 \Phi_c \Phi_c; \ 
d\cL = \fr12 K_{abc} d\L_{abc} = K_{abc} d(h_{a\n,\l} h_b^{\ \n} h_c^{\ \l});
\]
where $K_{abc} = \L_{abc} + \g S_{abc} - 
\s(\eta_{ab} \Phi_c - \eta_{ac} \Phi_b )$. 
Varying $h{\cal L}$, we  separate the
skew-symmetric part:
\begin{equation}\label{eB}
\begin{array}{l}
\bB_{a\m} = - h^{-1} g_{\m\n}
 {\displaystyle \frac{\d (h\cL)}{\d h^a{}_\n } } =
 K_{a\m\n;\n} + \L_{bca} K_{bc\m} - h_{a\m} \cL =0 \
( \bB_{a\m;\m} - \bB_{bc} \L_{bca} \equiv 0);
\\[4.5mm]
\bB_{[\m\n]}  = (\fr12 + \g)(S_{\m\n\l;\l} -
 \L_{\m\ve\t} \L_{\ve\t\n} + \L_{\n\ve\t} \L_{\ve\t\m} ) 
+ \fr{1-\s}2 (f_{\m\n} - \Phi_\l \L_{\l\m\n}) =0.
\end{array}
\end{equation}
The case $\s=1, \ \g=-\fr12$ corresponds to the vacuum equation 
of General Relativity 
\[ R_{\m\n} - \fr12 g_{\m\n}R =0; \]
 $2\cL$ coincides with 
the scalar curvature $R$ up to the $\Phi_{\mu;\mu}$ term
 [or up to $(h\Phi^\m)_{,\mu}$ 
for the density $h\cL$].\footnote{Recently, $F(T)$-gravities 
(where $T\sim\L^2$ is
the Ricci scalar minus $\Phi_{\m;\m}$ term) gained some 
popularity, e.g.\ see the {\em weak } $F(T)=T + a T^2$ gravity in
\href{http://arxiv.org/abs/2104.00116}{arXiv:~2104.00116}. However, the equations of these gravities, at least their skew-symmetric parts (having the form $\L^2 \L'=0$ for the {\em weak} $F(T)$-gravity), are evidently
nonregular [sure if $F(T)\neq T$].}

If $\s=1, \g\neq - 1/2$, the prolongation $\bB_{[\m\n];\n} =0$ results 
in a nonregular second order equation; while if $\s\neq1, \g=-1/2$,
such an equation, $\L \L' =0$, follows from $\bB_{[\m\n;\l]} = 0$ 
[see the last of Eqs.~\eqref{eB}].
These two parameter sets are inadmissible ("pathological") in Eq.~\eqref{eA} too.

The two-parameter class I$_{221}$ (Eqs.~(11') and (11a') in \cite{EiMa}) is possible, i.e.\
satisfies the identity \eqref{id8}, only for $D=4$ (apart from the particular case when Eq.~\eqref{e6S} coincides with the GR vacuum equation), 
and in this case $J_\mu =0$, but the conditions (\ref{tri-c}) cannot be 
totally satisfied.

The last and the most interesting class II$_{221221}$ 
(with one parameter $\s$) can be written as
follows:
\be \label{eD}
\fr12 (\bG_{a\m} - \bH_{a\m}) = L_{a\m\n;\n} - 
(1-2\s) (f_{a\m} + L_{a\m\n} \Phi_\n) =0,
\ee
where $L_{a\m\n} = \L_{a\m\n} - S_{a\m\n} - 
\s (h_{a\m}\Phi_\n - h_{a\n}\Phi_\m)$; all the conditions (\ref{tri-c}) are
satisfied.
Even though this one-parameter class of equations 
is not explicitly written in \cite{EiMa}, Eq.~(\ref{eD}) 
can be constructed from the Table (when doing so, one has to exchange 
$a_1$ and $a_2$ in Eq.~(1) -- one can just compare Eqs.~(1) and (6) 
of \cite{EiMa}; the parameter $q$ is related to $\s$ by
$\s = q/(2q -1)$; for variants $q= \pm 1$, the approach of \cite{EiMa}
is, strictly speaking, inappropriate).

For $\s= 1/2$, Eq.~(\ref{eD}) intersects with the class (\ref{eA});
for $\s=1/3$ (when $\t=0$), it coincides with the system from \cite{me_T}.
Also interesting is the case $\s=1$, when the symmetric part can be represented as a GR equation with "matter":
\[ -\fr12\bG_{\m\n} = R_{\m\n} - \fr12 g_{\m\n} R - T_{\m\n}(\L^2)=0
\ \, (R_{a\m\n\l} = 2h_{a\m;[;\n;\l]}\, , \ \, R_{\m\n} = R_{\ve\m\ve\n});
\]
however, $T_{\m\n}$ does not satisfy the weak energy condition \cite{HawEll}.

The identity (\ref{id8}) and the Maxwell equation for the class (\ref{eD})
have the form
\be \label{eM}
\begin{array}{c}
\t \bG_{a\m;\m} + (1-\s)\bH_{a\m;\m} \equiv (1- 2\s)
[\L_{a\n\l}\bH_{\n\l} + \t (\bG_{a\m} + \bH_{a\m})\Phi_\m],
 \ \t=1-3\s;  \\[2.5mm]
(f_{a\m} + L_{a\m\n} \Phi_\n)_{;\m} =0, \textrm{ or } 
(f_{\m\n} - S_{\m\n\l} \Phi_\l)_{;\n} =0.
\end{array}
\ee

There is a special space-time dimension $D=D_0$, where 
$D_0 = 1 + 1/\s$, for which the trace part of the system
loses its principal term and becomes 
nonregular (if $D_0$ is integer):
\[  \bG_{\m\m} = 2\s(D_0 - D) \Phi_{\m;\m} + Q_{\m\m}(\L^2) =0
 \ ( Q_{\m\m} \nequiv 0 )  .
\]
The linearized equations (\ref{e6S}), (\ref{e7A}) for weak fields
$h_{a\m} = \eta_{a\m}+\ve_{a\m}$
 at $D=D_0$ (when the trace equation vanishes, but one still can add
$\Phi_{\mu,\m}=0$) are invariant under infinitesimal conformal transformations
(with $\Phi_\m$ undergoing a gradient transformation):
\[ \ve^*_{a\m} = \ve_{a\m} + \eta_{a\m} \l(x^\n), \
\Phi^*_\m = \Phi_\m + (D-1) \l_{,\m} . \]

We are most interested by the case (\ref{eD}) 
with $\s= 1/3$ and $D > D_0=4$ (say, $D=5$, having in mind that the 
additional dimension undergo some kind of 
reduction).\footnote{Because this case, perhaps, is free from singularities 
in solutions of general position, see
\href{http://www.arxiv.org/abs/gr-qc/0412130}{%
arXiv: gr-qc/0412130v2}.
} 

Equations (\ref{eD}) include all cases with $a_2=a_3=0, \ J_\m\neq 0$,
and after the substitution
\be \label{sub}
h_a{}^\m = H^{-p} H_a{}^\m, \ \textrm{where } 
p=1/(D_0-D), \ H =\det H^a_{\;\m} \ (H_a{}^\m = h^{\s/(1+\s)} h_a{}^\m ),
\ee
the equation $H^{3p} (\bH_a{}^\m - \bG_a{}^\m)/2 =0$
takes a ``tri-linear'' form (it includes only the matrix $H_a{}^\m$
and its derivatives, but not 
$H^{a}_{\;\m}$; $H^{a}_{\;\m} H_a{}^\n = \d_\m^\n$):
\[ \hspace*{25 mm}
 [N^\m{}_{\!\!\!\!ab} H_b{}^\n - 
(H_a{}^\n{}_{\!\!,\,b} - H_b{}^\n{}_{\!\!,\,a})H_b{}^\m ]_{,\,\n} +
(1-2\s) [ H_b{}^\m (C_{a,\,b} - C_{b,\,a} ) - N^\m{}_{\!\!\!\!ab} C_b ] =0,
\hspace{21 mm}  (12')
 \] 
where $\ C_a = H_a{}^\l{}_{\!\!,\,\l} \,; \ \
{}_{,\,a} = {}_{,\,\l} H_a{}^\l \, ; \  \
N^\m{}_{\!\!\!\!ab} = H_a{}^\m{}_{\!\!,\,b} - H_b{}^\m{}_{\!\!,\,a}
+ \s (H_a{}^\m C_b - \eta_{ab} C_d H_d{}^\m) \, $ --
linear and bilinear terms.

Equations (\ref{eD}') remains regular for degenerate (but finite)
matrices $H_a{}^\m$ if $r = \textrm{rank\,} H_a{}^\m \geq 2$ (see \cite{me_3}).
One can construct a formal solution as a series, starting with a 
degenerate matrix $H_a{}^\m(x^\n_{(0)})$.

The substitution $h_a{}^\m \to H_a{}^\m$, which does not allow the inverse
map when $D=D_0$, ``corrects'' for the nonregularity of Eq.~(\ref{eD})
 in this case.

4. For a some given global rotation (\ref{e1}), a spherically symmetric
(or $O_{D-1}$-symmetric) field
$h^a{}_{\!\m}$ can be written as follows (see \cite{Ei_2}); 
$i,j,k = 1,\ldots,D-1; \ x^0 = t$):
\begin{equation}
\label{spsy} h^{a}{}_{\mu }(t, x^i)= 
\begin{pmatrix}
a & bn_{i}  \cr
cn_{i} &en_{i}n_{j}+d\Delta _{ij} 
\end{pmatrix}
 , \ \ \ h_{a}{}^{\mu }=\frac 1 \varkappa 
\begin{pmatrix}
e & -cn_{i} \cr
 -bn_i & a n_{i}n_{j}+\frac \vk  d \Delta_{ij}
\end{pmatrix} , \ \ \vk = ae - bc.
\end{equation}
Here $n_i = x^i/x$ is a unit vector along the radius $x = \sqrt{x^i x^i}$,
$\Delta _{ij}=\delta _{ij}-n_{i}n_{j}$ is the
 tensor orthogonal  to
 $n_{i}$, $\Delta_{ii}=D-2=k$ ($n_{i,j}=\Delta _{ij}/x$);
 $a,\ldots
,e$ are smooth functions of radius and time
 $t=x^{0}$ ($a,d,e$ are even, while
  $b$ and $c$ are odd functions of radius $x$);
 $ h = \det h^{\,a}{}_\m = \vk\,d^{\,k}$.

The co-frame and contra-frame in (\ref{spsy}) are invariant under the next coordinate transformations:
\[ x^* = X(x,t), \ t^* = T(x, t); \  \textrm{these keep } n^*_i = n_i.
\]
Notice that $d$ transforms independently: $d^* =d x/x^*$;
that is, the product $u=xd$ is a 
scalar.\footnote{ In principle, one could choose it
as the new radius, $x^*=u$ (then $d^*=1$); later we'll choose 
a bit different scalar.}

For the field (\ref{spsy}) it is clear that $S_{\m\n\l}\equiv 0$
 (no ways to build this skew-symmetric tensor),
and so we obtain, from the last equation in (\ref{eM}),
 two easily integrable equations
for $\omega = h f^{0i} n_i$ ($f^{ij} \equiv 0$):
\be \label{i1}
 (hf^{\mu \nu })_{,\nu }= 0 ;\ \Rightarrow (\omega n_{i})_{,i}
= \om^\prime  + \om k/x =0 ,\ \  
\dot{\om}=0 ;\ \textrm{ hence }  \ \om=C/ x^{k} .
\ee
The "dot"  denotes differentiation by time, and
 the "prime" -- with respect to the radius.
A  $C\neq 0$ (the integration constant) clearly means, in fact, adding to Eq.~(\ref{eM}) a $\delta$-source (to the current); in order to avoid modifying the equations, we should choose  $C=0$. So $f_{\m\n}=0$ -- \,this follows 
immediately from the antisymmetric part if $\s \neq 1/3$ (i.e.\ $\t\neq 0$).

Taking into account that $S_{\m\n\l} =0, \ f_{\m\n} = 0$, we can rewrite 
Eq.~(\ref{eD}) in the following form:
\be \label{eE}
(hL_a{}^{\m\n})_{,\n} = (1 -2\s)hL_a{}^{\m\n}\Phi_\n , 
\textrm{\, or } E_a{}^\m =(h\psi^{2\s-1} L_a{}^{\m\n})_{,\n} =0;
\ee
here $L_{a\m\n} = \L_{a\m\n} - \s(h_{a\m}\Phi_\n - h_{a\n}\Phi_\m)$
and we introduce the scalar $\psi$: $\Phi_\m = \psi_{,\m}/\psi$.

The equations $E_{\underline{0}}{}^0 = 0,\, E_{\underline{0}}{}^i = 0 $
can be integrated in the same way as Eq.~(\ref{i1}). Choosing the 
integration constant from the condition that $\d(x)$ terms vanish,
we obtain $L_{\underline{0}}{}^{0i} = 0 = L_{\underline{0}0i}$
($L_{\underline{0}}{}^{ij} \equiv 0$). Underlined indices represent particular
values (scalar, or {\em inertial}) of Roman indices, so $\Phi_{\underline{0}}\neq\Phi_0$.

The equation $n_i L_{\underline{0}0i} = 0$ looks as follows
\[ n_i \left(h_{\underline{0}0,i} - h_{\underline{0}i,0} 
-\s h_{\underline{0}0} \psi_{,i}/\psi +\s h_{\underline{0}i} \psi_{,0}/\psi
\right) = -a' + \dot{b\,\,}\! +\s\, a\, \psi'\!/\psi - 
\s\, b\, \dot{\psi\,\,\,}\!\!\!/\psi =0;
\]
  this can be also written as a conservation law, so one can introduce
a scalar $\tau$:
\[ (b\,\psi^{-\s}\dot{)\,}\! - (a\,\psi^{-\s})' =0; \
d\tau = a\psi^{-\s} \, dt + b\psi^{-\s}\, dx\,; \
\dot{\t\,\,}\!\!=a\psi^{-\s}, \ \t' = b\psi^{-\s}, \ \dot{\t\,\,\,\,}\!\!\!'  
=\dot{\t\,\,\,\,\,\,\,} \!\!\!\!\!\!\! ' \,.  
\]
This scalar can be chosen as the new 
time,\footnote{In a similar way, for trivial solutions with
$\L_{a\m\n}=0$, one can introduce the scalars $y_a$ through
solving eq-s $y_{a,\m} = h_{a\m}$ 
($y_{a,\m \n} - y_{a,\n \m} = 0 = h_{a\m,\n} - h_{a\n,\m} $); 
the scalars $y_a$ can serve as  preferable 
coordinates, inertial ones, where $h^a{}_\m = \d^a_\m$.}
$t^*=\tau$; with this 
coordinate choice we have
\[ b=0\,, \ a= \psi^\s\,. \]

For the components $\L_{a\m\n} $ and $\Phi_\m = \psi_{,\m}/\psi$,
 we obtain the
expressions [see Eqs.~(\ref{e2}), (\ref{spsy}); $b=0$]:
\be \label{eL}
\begin{array}{c}
\L_{\underline{0}0i} =- a' n_i, \
\L_{\underline{i}j0} =\! \dot{\,e}\, n_i n_j + \dot{\,d}\, \D_{ij}
- c' n_i n_j - c \D_{ij}/x , \ 
\L_{\underline{i}jk} = \left[ d' - (d-e)/x \right]
(\D_{ij} n_k -\D_{ik} n_j) ; \\[2.5 mm]
\Phi_0 = h_{\underline{i}}{}^j \L_{\underline{i}j0} = 
(\!\dot{\,e}- c')/e + k\,(\!\dot{\,d}-c/x)/d + ca'/(ae)\, , \
\Phi_k n_k 
 = a'/a + k \left[ d' + (d-e)/x \right]/d \, .
\end{array}
\ee

Now let us write the non-zero components of $L_{a\m\n}$ 
(using $\s \Phi_\m = a_{,\mu}/a $):
\[  \hspace*{6 mm}  
L_{\underline{i}j0} = (\! \dot{\,e} - c' - e\dot{\,a}/a +c\,a'\!/a)
\, n_i n_j  +
(\!\dot{\,d} - d \dot{\,a}/a)\, \D_{ij} , \ 
L_{\underline{i}jk} = \left[ d' + (d-e)/x - d\, a'\!/a\right]
(\D_{ij} n_k - \D_{ik} n_j)\, . \hspace{5 mm}  (18')
\]

To further fix the coordinate freedom, one can choose  $c =0$.
 This restriction 
results in a first order equation for $X(x,t)$, i.e., there remains a freedom in 
the choice of a function of one variable ("initial condition").
In what follows, we shall have to choose one
integration function, which will be done taking into account convenience and
the boundary condition: $h^a{}_{\!\m} \to \d_a^\m $ as $x \to \infty$.

Integrating $\Phi_0 = \dot{\psi\,\,\,}\!\!\!/\psi$ [full time derivative, see
$\Phi_0$ in Eq.~(\ref{eL})], we obtain $\psi = e\,d^k$.

Now it is time to go over to the components $h \psi^{2\s-1} L_a{}^{\m\n} $
[see Eq.~(\ref{eE}); $h = a\,e\,d^{\,k} =a\,\psi, \ a=\psi^\s$]:
\be \label{eLL}
a^3 L_{\underline{i}}{}^{j0} = \dot{\,\a}\, n_i n_j + \dot{\,\b}\,\D_{ij}\, ,
 \ a^3 L_{\underline{i}}{}^{jk} = [-\a^2 \b' +\a\b (\a - \b)/x ]
(\D_{ij} n_k - \D_{ik} n_j)\, ;
\ee
here we have introduced new variables 
$\a =a/e = d^{\,k\s}e^{\,\s -1}, \ \b = a/d$.

The mapping $e,d \to \a,\b$ (non-invertible at $ k= 1/\s -1$)
corresponds to the change of variables (\ref{sub});
we also write out $g^{\m\n}$:
\[  H_a{}^\m = 
\begin{pmatrix}
1 & 0 \cr 
0 & \a\, n_i n_j + \b\, \D_{ij}
\end{pmatrix}
= h^{\s/(1+\s)} h_a{}^\m = a\,h_a{}^\m  \,; \ \,
g^{\m\n} = 
\begin{pmatrix}
-1/a^2 & 0 \cr 
0 & n_i n_j/e^2 + \D_{ij}/d^2
\end{pmatrix}
\, .  \]
Substituting Eq.~(\ref{eLL}) in Eq.~(\ref{eE}), we can write the equation
$E_{\underline{i}}{}^j n_i n_j =0$ ($L_{\underline{i}}{}^{jk} n_i \equiv 0$):
\be \label{e20}
(a^3 L_{\underline{i}}{}^{j0}\dot{)\,\,}\! n_i n_j 
 - a^3  L_{\underline{i}}{}^{jk}  n_{i,k}\, n_j =
\ddot{\a} - k\,[\a^2\b' -\a\b(\a-\b)/x ]/x = 0.
\ee

The equation $E_{\underline{i}}{}^0 n_i  =0  $ becomes just 
\[ ( a^3  L_{\underline{i}}{}^{0j} )_{,j} \,  n_i =
( a^3  L_{\underline{i}}{}^{0j}    n_i)_{,j} -
a^3  L_{\underline{i}}{}^{0j}\, \D_{ij}/x =0 
= [-\a' - k\,(\a - \b)/x\dot{]\,\,\,\,} \!\! ; 
\textrm { \ hence, }\, \b = \a + x\,\a'/k \, ,
\]
 which is a simple constraint equation,\footnote{The same follows from the eq-n
$\Phi_i n_i = \psi'/\psi = e'/e + k\,d'/d$, see (\ref{eL}).}
 and after its substitution into Eq.~(\ref{e20}) one  obtain ($k>0$)
\be \label{e21}
\ddot{\a} = \a^2 \a'' + \a\, \a'^2/k + (k+2)\,\a^2 \a'/x\,;
 \ \,  \b = \a + x\,\a'/k\,.
\ee
The equation  $E_{\underline{i}}{}^j \D_{ij}  =0 $ agrees with the system (\ref{e21}).

The Cauchy problem is reduced to solving one second order equation; one can 
take the initial conditions to be even functions
$\a_0(x) > 0$ and $\dot{\a\,}_{\!0}(x)$.
 These functions can be defined so that
$\b_0(x)$ vanishes in a point $x_1 >0$, but $\b(x,t) >0$ when $t<0$ 
($\a_0 >0$ in the neighborhood of $x_1$), e.g.
\[ \b_0(x) =  \b(x,t=0) = 1 - 3x^{2k}/(2+x^{6k})\geq 0, \ 
\min \b(x) = \b(x_1 = 1)= 0;
 \ \a_0(x) = x^{-k} \!\! \varint\! \b_0(x)\,dx^k >0.
\]
At the point $x = x_1 =1, t=0$ there is singularity, since $\b_0(1)=0$;
when $t>0$, a region appears in which $\b$ is negative (if $\dot{\b}_0(1)<0$).
If one allows the initial conditions to contain a region in which $\b<0$,
then they can be defined so that $\a_0$ vanishes (touches zero)
at one point $x_2$, $\b_0$ changing sign, and for $t>0$, a region appears with
$\a<0$ (if $\dot{a}_0 < 0$ in a neighborhood of $x_2$).

The points with $\a=0$ or $\b=0$ are singular when $D>D_0$ (and $D<D_0-1$) or
$(D-2)/(D_0-D)<1$ because $\Phi_{\underline{0}}$ becomes infinite:
\[ \Phi_{\underline{0}} = -p/\s (\dot{\a}/\a + k \dot{\b}/\b )\a^p \b^{kp}; \
 p = (D_0 - D)^{-1}, \ D_0 = 1 + \s^{-1}. \]
It seems, all combinations of signs of $\a$ and $\b$ are 
possible. 

Still strong doubts exist that this coordinate choice, which
`diagonalizes`  the frame field ($b=0$ and also $c=0$), is available for all $O(n-1)$-symmetrical
solutions (i.e. for spherically symmetrical solutions of general position).

5. Let us consider another way, perhaps more reasonable (as it is of more
covariant sense), to set the second 
condition on the coordinates (on the radius $x$). We can rewrite the
equation $\Phi_i n_i = \psi'/\psi$ in the next form [see the last equation
in (\ref{eL}); $b=0$  and $a=\psi^\s$]:
\[ (xd)'  - e  = \frac{1-\s}k \, xd \, \psi'/\psi, \
\textrm{or }  (\psi^{-w} \, xd)' = \psi^{-w}\, e, 
\textrm{ where } w = \frac{1-\s}k.  \]  
Let us take the 
scalar\footnote{One should remember that the product 
$xd$ behaves as a scalar.} 
$\psi^{-w} xd$
as the new radius $x^*$; with this second restriction on the 
coordinate choice we obtain ($h = a\,e\, d^k; \ a = \psi^\s$)
\[ \psi^{-w} d = 1 = \psi^{-w} \, e, \, \textrm{ i.e. }
e = d = \psi^{(1-\s)/k} ;
 \ \, d^k = \psi^{1- \s}, \ \, h\,\psi^{2\s-1} = a^2 e \, . \]
Using this equals,  we rewrite the equation 
$\Phi_0 - \dot{\psi\,\,}\!\!/\psi =0$ [see (\ref{eL}); $d^k =\psi^{1-\s}, \ a=\psi^\s, \ d=e$]: 
\[   
\dot{e\,}\!/e  + k \dot{d\,}/d - \dot{\psi\,\,}\!\!/\psi 
- c'/e - kc/(ex) + ca'/(ae)= 0 = (\dot{e\,}\!/e - \dot{a}/a)
- (c/e)' - k(c/e)/x + (a'/a - e'/e)c/e, \]
\[ \textrm{ i.e. } 
 \dot{A\,\,}\!\!/A + BA'/A = B' + kB/x, \textrm{ where we use } \,
B = -c/e, \ A = a/e =  e^{k\s/(1-\s) -1}; \ \,
\ h_a{}^\m = a^{-1}
\begin{pmatrix}
1 & B n_i \cr 
0 & A \d_{ij}
\end{pmatrix}
.
\]
The components 
of $h \psi^{2\s-1} L_a{}^{\m\n}$ now look simple
[$g^{\m\n} = \frac1{a^2 e}
\begin{pmatrix}
-e & c n_i \cr 
cn_i & \frac{a^2 {-} c^2}e  n_i n_j + \frac{a^2}e \D_{ij}
\end{pmatrix} $; see (\ref{eL})', (\ref{eLL})]:
\[ 
a^2 e\, L_{\underline{i}}{}^{j0} =
(k\, B /x )n_i n_j
+ [ B' +(k-1)B/x]\D_{ij}\, , \ \,
a^2 e\, L_{\underline{i}}{}^{jk} = [B B' +(k - 1)B^2/x - AA'] (\D_{ij} n_k - \D_{ik} n_j)\,  .
\] 

The equation $E_{\underline{i}}{}^j n_i n_j =0 =  
(a^2e\, L_{\underline{i}}{}^{j0}\dot{)\,\,}\! n_i n_j 
 - a^2 e\,  L_{\underline{i}}{}^{jk}  n_j\, \D_{ik}/x $
has the following form
\[
 (kB/x \dot{)} + k[B B' - A A' +(k-1)B^2/x ]/x = 0\, , \]
and it gives the other first order equation (after removing the
factor $k/x$), so in the end we arrive to   
the next system of quasilinear equations (which closely resembles
the 2D equations of ideal gas dynamics for Chaplygin gas):
\begin{equation}\label{e22}
\begin{array}{l}
\dot{A\,\,\,}\!\!\! = A B' - B A' + k AB/x \, , \\[1.5 mm]
\dot{B\,\,}\!\! = A A' - B B' - (k - 1)B^2/x \, ,
\end{array}
\end{equation}

As initial conditions, one has to choose here a positive even function
$A_0(x)$ ($A_0 \to 1$ as $x \to \infty$) and an odd function $B_0(x)$.

Going over to the Riemann invariants $u=A+B, \ v=A-B$ it is easy to show
that the system (\ref{e22}) belongs to the weakly nonlinear type (see \cite{RoYa}), i.e., there is no gradient catastrophe.

Eq.~(\ref{e22}) has no stationary solutions apart from the trivial
solution $A =1,\ B=0$. During a finite time, a singularity
(where $1/A,\, 1/B$, or $A \to 0$) cannot occur in the solutions
of Eq.~(\ref{e22}).

The first equation in (\ref{e22}) can be written as a conservation law
\[  (x^k/A \dot{)\,\,}\! =-(x^k B/A)', 
\]
so that one can introduce a Lagrange variable $y(x,t)$
(see \cite{RoYa}; $x_y$ means $\partial x/\partial y$):
\[ dy = x^k A^{-1}\, dx - x^kB\,A^{-1}\, dt\, ; \
dx = Ax^{-k}\,dy + B\,dt\,,\ B = \dot{x}\, , \ A= x^k\,x_y\, .
\]
Substituting 
\[ A,\, B,\, \dot{B\,}(x,t) = \ddot{x} - \dot{x}{}_y \dot{x}/x_y, \, 
 B' = \dot{x}_y/x_y, \, A' \]
 in the second equation of (\ref{e22}),
after the change of variables $z = x^k$ we obtain:
\[ \ddot{z} = z^2 z_{yy} + z\,z^2/k \, . \]
At $k =1$, there exists a Lagrangian $L = \dot{z}^2 - z^2 z_y^2$
[and for Eq.~(\ref{e21}), $L = (\dot{\a}^2 - \a^2 \a'^2)x^3$].

6. The last second order equation matches Eq.~(\ref{e21}) 
with the last term, $\sim\a^2\a'/x$, removed; this term does not matter for
very high stages 
of expansion, when the spherical wave looks  almost like a plane longitudinal
wave. On the other hand, the system (\ref{e22}) without terms with radius,
$\sim 1/x$, becomes completely integrable (e.g., by the hodograph method) having 
infinite number of conservation laws.
Perhaps this hints that the spherically symmetrical problem [for the class
(\ref{eD})] admits exact solutions, as well as the problem with plane longitudinal waves.

This O${}_4$-symmetrical longitudinal single wave can serve, for the case $\t=0$
($\s=1/3$) and $D=5$, as a valid cosmological background, a sort of expanding
$S^3$-shell, or brane, working like a waveguide for ensembles of shorter waves of different polarizations and also for topological quasi-particles.
This ultra-relativistically expanding cosmology leads to the simple Hubble diagram \cite{more}:
\[ \mu(z)=\mu_0+5\log[(1+z)\,\ln(1+z)] \
(\textrm{where } \mu_0 = -5\log( H_0\,d_*/c)\approx 43.3, \ d_* = 10\,\textrm{pc}).
\] 

The question about singularities in the spherically symmetric solutions still
 needs a more careful consideration. One could write 2D-covariant 
(arbitrary radius and time)
 system of equations for  the scalar ($u=xd$) and two vectors (or one 2D-frame)
for general coordinates (no gauge restrictions). 
Then, the compatibility test of this
2D-covariant system should be extended on the cases where the
2D co-frame (or its inverse, i.e.\ contra-frame density of some weight)
becomes degenerate, or where 
the scalar becomes zero or infinite.\footnote{Such a covariant
approach (for solutions of general position, not symmetrical) is
used in \href{http://www.arxiv.org/abs/gr-qc/0412130}{%
arXiv: gr-qc/0412130v2}. }
\\[.1 mm]

The author thanks I.\,L.~Bukhbinder for his interest to this paper and useful 
discussions during the seminar. Thanks also go to A.\,V.~Shemyakin who had kindly sent author hard copies of some papers (see v.~1 of this paper).

\end{document}